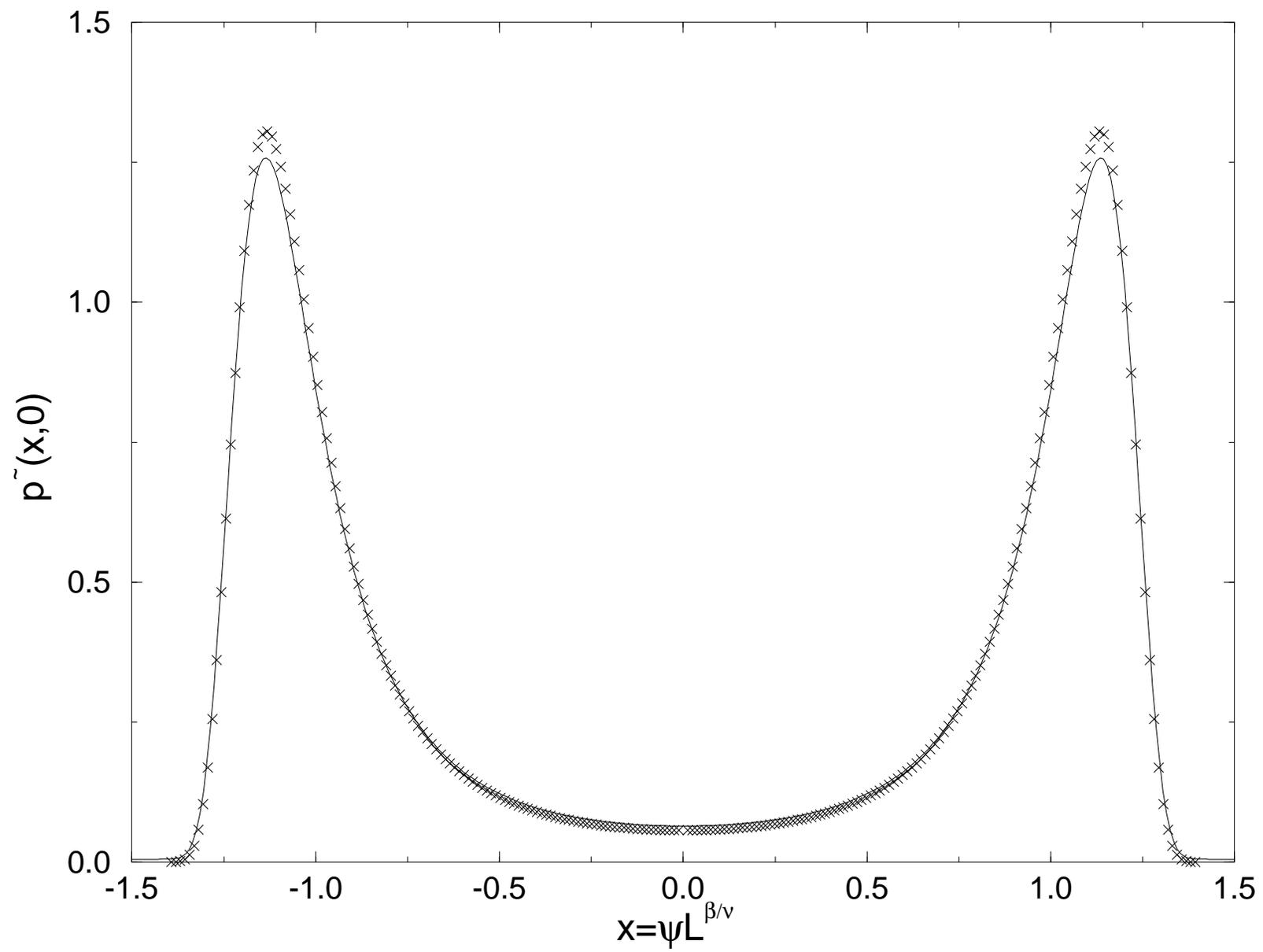

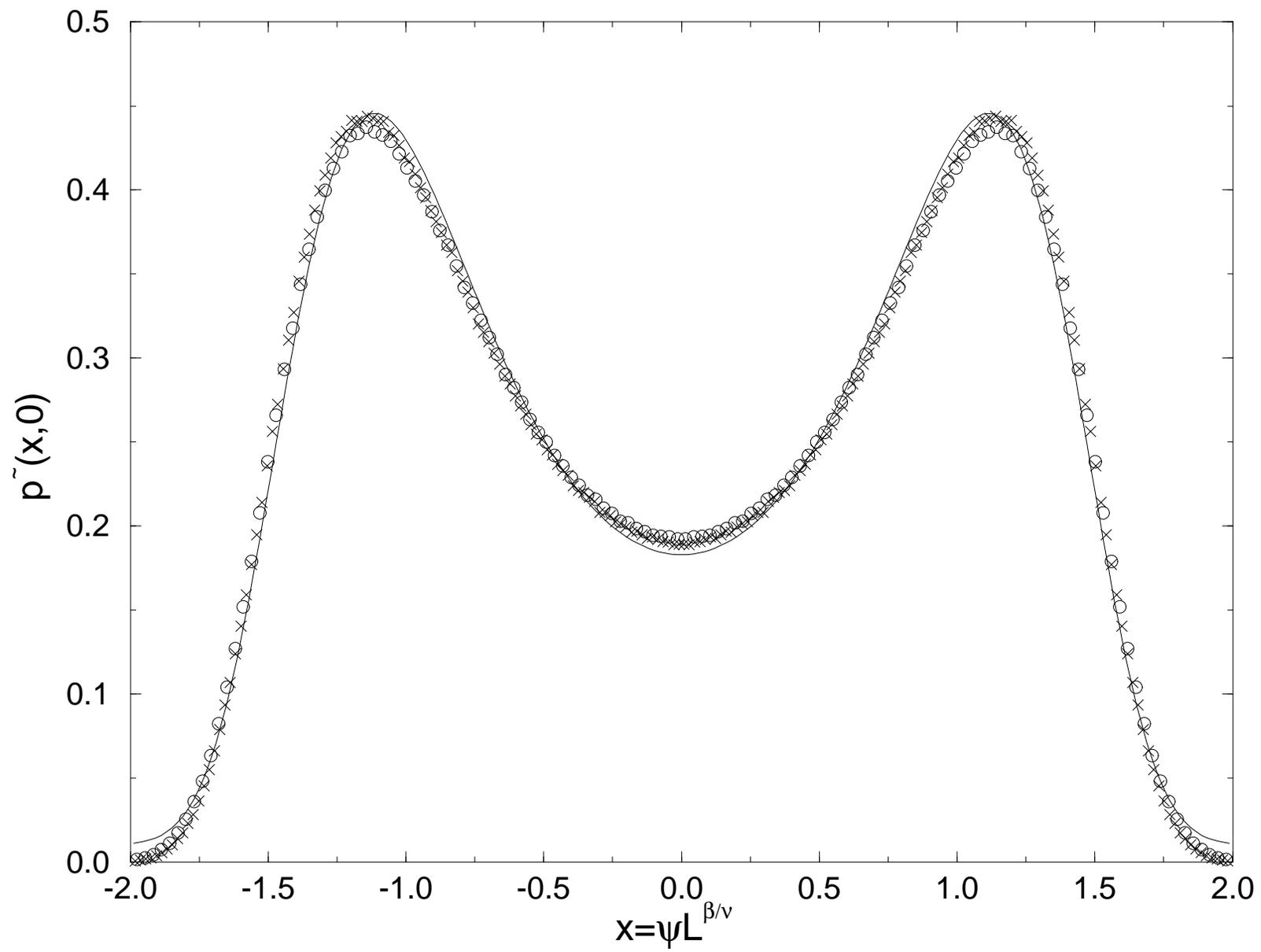

# ARE CRITICAL FINITE SIZE SCALING FUNCTIONS CALCULABLE FROM KNOWLEDGE OF AN APPROPRIATE CRITICAL EXPONENT ?

R. Hilfer[1,2] and N.B. Wilding[1]

[1] *Institut für Physik, Universität Mainz, 55099 Mainz, Germany*
[2] *International School for Advanced Studies, Via Beirut 2-4, 34013 Trieste, Italy*

## Abstract

Critical finite size scaling functions for the order parameter distribution of the two and three dimensional Ising model are investigated. Within a recently introduced classification theory of phase transitions the universal part of critical finite size scaling functions has been derived by employing a scaling limit which differs from the traditional finite size scaling limit. In this paper the analytical predictions are compared with Monte Carlo simulation results. We find good agreement between the analytical expression and the simulation results. The agreement is consistent with the possibility that the functional form of the critical finite size scaling function for the order parameter distribution is determined uniquely by only a few universal parameters, most notably the equation of state exponent.

A universality class in the theory of critical phenomena is identified by a set of critical exponents and a set of universal scaling functions [1–3]. In practice finite size scaling functions and their associated universal amplitudes or amplitude ratios have become an indispensable tool for the extraction of universal behaviour from numerical simulations of finite systems throughout many fields of physics [4–8], and the subject remains of vigorous research interest [9–12]. It is therefore of broad interest to obtain exact information on finite size scaling functions. Such knowledge would also be particularly important for discerning corrections to scaling behaviour.

Most determinations of critical finite size scaling functions or amplitude ratios have been carried out by numerical simulation of critical systems [3,13–21]. Analytical calculations [22–25] are complicated because at criticality the influence of boundary conditions cannot be neglected. This renders renormalization group methods difficult to implement. The problem of hyperscaling violations [26,27] has further obscured the basic question as to what extent finite size scaling functions are universal or not. Recently [28], however, finite size scaling theory has been reanalysed from the perspective of a general classification theory for phase transitions [29–33]. In these papers it was shown that phase transitions in statistical mechanics may be classified according to a generalized classification theory. This theory classifies each transition according to its generalized order in analogy with Ehrenfest's classification scheme. The objective of the present letter is to compare analytical predictions of the generalized classification scheme with computer simulation results. The comparison is carried out for the critical finite size scaling function of the order parameter distribution in the two- and three dimensional Ising models with periodic boundary conditions. Good agreement between simulations and the theoretical predictions is found over the entire range for which numerical data are available.

Define $p(\psi, L, \xi)$ to be the probability density function for the fluctuating order parameter $\Psi$ in a finite system of size $L$ and order parameter correlation length $\xi$. Then the scaling function of interest $\widetilde{p}(x,y)$ is defined by

$$p(\psi, L, \xi) = L^{d(d_\Psi - d^*)/(d-d^*)} \, \widetilde{p}(\psi L^{d(d_\Psi - d^*)/(d-d^*)}, L/\xi_{d^*}) \tag{1}$$

where $d_\Psi$ is the anomalous or scaling dimension of the order parameter, $d^*$ is Fishers anomalous dimension of the vacuum [34], and $\xi_{d^*}$ is Binders thermodynamic length [35]. If hyperscaling holds then $d^* = 0$, the thermodynamic length becomes the correlation length, $\xi_0 = \xi$, and the exponent in (1) reduces to the familiar form $d_\Psi = \beta/\nu$ where $\beta$ is the order parameter exponent and $\nu$ the correlation length exponent. The scaling function $\widetilde{p}(x,y)$ is expected to be universal up to the choice of boundary conditions [13].

Given the scaling Ansatz (1) the traditional scaling analysis [13,25] of the critical scaling function $\widetilde{p}(x,0)$ distinguishes two cases. (i) For $x \ll 1$ the scaling function is expected to have the universal Landau-Ginzburg form [13]

$$\widetilde{p}(x,0) \propto \exp(-A_0 - A_2 x^2 - A_4 x^4 - ...). \tag{2}$$

(ii) For $x \gg 1$ and groundstate boundary conditions (e.g. all spins positive for the Ising model) the scaling function is expected to have the squeezed exponential form [36]

$$\widetilde{p}(x,0) \propto \exp(-A x^{\delta+1}) \tag{3}$$

where $\delta$ denotes the equation of state exponent. Based on scaling arguments the same form is expected to apply for periodic boundary conditions [25].

Little exact information is available for $\widetilde{p}(x,0)$. To the best of our knowledge only the cumulant ratio $g(0) = (\int |x|^4 \widetilde{p}(x,0)\,dx)/(\int |x|^2 \widetilde{p}(x,0)\,dx)^2$ for the two dimensional Ising model with *singular periodic* boundary conditions has been calculated exactly [24]. For noncritical systems on the other hand the analogous noncritical scaling function is Gaussian by virtue of the central limit theorem [13]. The absence of exact information about the critical function $\widetilde{p}(x, 0)$ even for the otherwise exactly solvable two dimensional Ising model is related to the absence of its solution in nonzero magnetic field [36].

Recently the universal part of critical finite size scaling functions has been related to finite ensemble scaling functions [33,28]. Finite ensemble scaling functions arise in the ensemble limit while finite size scaling functions arise in the finite size scaling limit. For a $d$-dimensional discretized lattice system in the fully finite hypercubic geometry the finite size scaling limit is defined as the limit $\lim_{\substack{L,\xi\to\infty \\ L/\xi=c}}$ in which the box dimension $L$ and the correlation length $\xi$ increase to infinity in such a way that their ratio remains constant. In the ensemble limit on the other hand the lattice constant $a$ approaches 0 simultaneously. More precisely, the ensemble limit is defined as the limit $\lim_{\substack{M,N\to\infty \\ N/M=c}}$ in which $M = (\xi/a)^d$ and $N = (L/\xi)^d$ approach infinity such that their ratio remains constant.

In the finite size scaling limit the critical finite size scaling functions are found to contain a universal as well as a nonuniversal part. The universal part is given by the finite ensemble scaling functions which arise in the finite ensemble limit and for which analytical expressions can be derived if the scaling dimension of the critical operator in question is known. For a critical operator $X$ at the critical point of a $d$-dimensional system with Ising symmetry and periodic boundary conditions the universal part of the critical finite size scaling function is written as [28]

$$\widetilde{p}(x, 0) = \frac{1}{2}h^+(x;\varpi_X) + \frac{1}{2}h^-(x;\varpi_X) \tag{4}$$

with

$$\varpi_X = \min(2, 2 - \alpha_X) \tag{5}$$

where $\alpha_X$ is the thermodynamic fluctuation exponent [34] of the observable $X$. If $X$ represents the energy density then $\alpha_{\mathcal{E}} = \alpha$, the specific heat exponent, while for the order parameter density $\alpha_\Psi = 1 - (1/\delta)$ where $\delta$ is the equation of state exponent. The scaling functions $h^\pm(x; \varpi_X)$ obey $h^+(x) = h^-(-x)$ and can be written in terms of the $H$-function representation of stable probability densities [37–39] as

$$h(x; \varpi_X) = \frac{1}{\varpi_X} H_{11}^{10}\left(x \,\middle|\, \begin{array}{c}(1 - 1/\varpi_X, 1/\varpi_X) \\ (0, 1)\end{array}\right). \tag{6}$$

The general class of $H$-functions is usually defined in terms of Mellin-Barnes contour integrals and contains Meijer's $G$-function as well as many other generalized hypergeometric functions as special cases. For a precise definition we refer the reader to standard tables [38]. Note that equation (6) depends only on $\varpi_X$ which is completely determined by the scaling dimension of $X$. Note also that equations (4) and (6) apply for periodic boundary conditions. The general theory identifies a universal shape and symmetry parameter which is related to different choices of boundary conditions [28]. It was identified in [28] to be unity for periodic boundary conditions and the same choice has been applied here. In the following we focus on order parameter fluctuations, i.e. $X = \Psi$. In that case the index $\varpi_X$ becomes $\varpi_\Psi = 1 + (1/\delta)$ where $\delta$ is the equation of state exponent.

The scaling function given by (4) and (6) are consistent with the scaling results (2) and (3). The functions $h^\pm(x; \varpi_X)$ are entire functions of $x$, and thus $\widetilde{p}(x, 0)$ may be expanded around $x = 0$ as assumed in (2). Secondly the asymptotic expansion of the $H$-functions [40] gives

$$h^\pm(x; \varpi_\Psi) \propto x^{(\delta-1)/2} \exp\left[-\frac{1}{\delta}\left(\frac{\delta x}{\delta + 1}\right)^{\delta+1}\right] \tag{7}$$

which is consistent with the scaling result (3) even though it is not derived in the same scaling limit [41].

In Figure 1a we compare the analytical result to simulation data for the two dimensional Ising model with periodic boundary conditions. In this case $\delta = 15$ and the critical tem-perature is known exactly. The simulation results are represented as crosses, the analytical result as a solid line. The crosses give a smoothed representation of scaled simulation data [19] for system size $L = 64$. All distributions are scaled to unit norm and variance. For the analytical curve this requires a cutoff which was chosen at a value close to the largest simulation data point.

In Figure 1b the same comparison is shown for the case of the $d = 3$ Ising model. In this case neither the critical temperature nor the equation of state exponent $\delta$ are known exactly. The data points represent original high precision Monte Carlo simulations in which systems of size $L = 20$ and $L = 32$ were studied for $50 \times 10^6$ Monte Carlo sweeps each using a vectorized code on a Cray YMP. The data for the system size $L = 20$ is represented by crosses, that for $L = 32$ by circles. We used the estimates [20] $J/(k_B T_c) = 0.2216595$ for the critical temperature and $\delta = 4.8$ for the equation of state exponent. Here $J$ denotes the Ising exchange coupling and $k_B$ is Boltzmann's constant.

The agreement between the analytical prediction and the simulation results in both two and three dimensions is gratifying. We attribute the small discrepancies, in part at least, to the low statistics in the tails of the scaling function. This view is supported by comparing the scaling function obtained in a previous small-scale study of the 3D Ising model [13], with that of figure 2. While both the scaling functions of [13] and those reported here exhibit excellent data collapse, the scaling functions in both cases are markedly different, demonstrating the presence of the nonuniversal part [28]. The good agreement between theory and simulations lends substantial support to the theoretical ideas from which the scaling functions derive. In particular it should be emphasized that the only difference between the solid lines in Figure 1a and 1b is the value of $\delta$. This suggests that the *full functional form* of the universal part of the critical finite size scaling function of the order parameter distribution could in simple cases be determined by a few universal parameters, most notably the equation of state exponent. Further numerical and analytical studies are, however, required to conclusively establish whether the proposed universal scaling functions agree fortuitously with the Monte Carlo data or are more generally correct.

## ACKNOWLEDGMENTS

The authors are grateful to A.D. Bruce for comments and making available the results of Monte Carlo studies on the two dimensional Ising model. One of us (R.H) thanks the Commission of the European Communities (ERBCHBGCT920180) for financial support.

## FIGURE CAPTIONS

FIG 1 : Comparison of the analytical predictions of equations (4)-(6) with Monte-Carlo simulation data for the magnetisation distribution of the two and three dimensional critical Ising model with periodic boundary conditions. In (a) the 2D data of [19] (crosses) is compared with the analytical prediction for $\delta = 15$ (full curve). Part (b) shows original 3D simulation data for $L = 20$ (crosses) and $L = 30$ (circles), collected at temperature $J/(k_B T_c) = 0.2216595$. The corresponding analytical curve corresponds to the estimate $\delta = 4.8$ [20]. All data have been scaled to unit norm and variance and statistical errors do not exceed the symbol sizes.